\documentclass[prd,aps,preprint,amsmath,nofootinbib,amssymb,eqsecnum,showkeys]{revtex4-1}
\usepackage{verbatim,graphics,graphicx,color,slashed,textcomp}
\usepackage{ulem} 
\usepackage[colorlinks=true,linkcolor=red,urlcolor=blue,citecolor=blue]{hyperref}

\graphicspath{{figures/}}

\usepackage[utf8]{inputenc}

\usepackage[colorlinks=true,linkcolor=red,urlcolor=blue,citecolor=blue]{hyperref}

\begin{document}

\title{Inflation in gauge theory of gravity with local scaling symmetry and quantum induced symmetry breaking}
\preprint{******}


\author{Yong Tang$^{a}$ and Yue-Liang Wu$^{b,c,d}$}
\affiliation{
	${}^a$Department of Physics, Faculty of Science, \\
	The University of Tokyo, Bunkyo-ku, Tokyo 113-0033, Japan\\
	${}^b$International Centre for Theoretical Physics Asia-Pacific, Beijing, China \\
	${}^c$Institute of Theoretical Physics, Chinese Academy of Sciences, Beijing 100190, China\\
	${}^d$University of Chinese Academy of Sciences, Beijing 100049, China}

\begin{abstract}
Motivated by the gauge theory of gravity with local scaling symmetry proposed recently in~\cite{Wu:2017urh, Wu:2015wwa}, we investigate whether the scalar field therein can be responsible for the inflation. We show that the classical theory would suffer from the difficulty that inflation can start but will never stop. We explore possible solutions by invoking the symmetry breaking through quantum effects. The effective potential of the scalar field is shown to have phenomenologically interesting forms to give viable inflation models. The predictions of physical observables agree well with current cosmological measurements and can be further tested in future experiments searching for primordial gravitational waves.

\end{abstract}

\maketitle

\section{Introduction}
Inflation has been one of the popular paradigms to solve several notable problems in cosmology since 1980s. It provides a basic framework to explain the origins of initial conditions in standard big-bang theory. In the inflationary epoch, a scalar field is usually dominating the energy density of the universe and results in an exponential expansion of the background spacetime. Quantum fluctuation of this scalar field is responsible for the primordial inhomogeneity and anisotropy that will lead to our observed cosmos.

Motivated by the gauge theory of gravity with local scaling symmetry proposed recently in~\cite{Wu:2017urh, Wu:2015wwa}, we investigate whether the scalar field therein can be responsible for the inflation. In Refs.~\cite{Wu:2017urh, Wu:2015wwa}, a general hyperunified field theory of gravity is constructed, incorporating the spin gauge group and local scaling symmetry. To make the theory scaling invariant, a fundamental real scalar field $\phi$ and its corresponding Weyl gauge field $W_\mu $ have to be introduced. Another basic field $\chi^a_{\ \mu}$ is related with the traditional metric tensor through $g_{\mu\nu}=\chi_{\ \mu}^{a}\chi_{\ \nu}^{b}\eta_{ab}$ where the sign convention $\eta_{ab}=(1,-1,-1,-1)$ is adopted.

Induced gravity can be generally referred as theories where the Planck scale is generated by other fields~\cite{Zee:1978wi, Adler:1982ri}. In the literature, global scaling symmetry has been used in building inflation models (see~\cite{Wetterich:1987fm, Rinaldi:2014gha, Kannike:2015apa, Ferreira:2018qss, Csaki:2014bua, Kaiser:1994vs, GarciaBellido:2011de, Kannike:2015kda, Salvio:2017xul, Karam:2017rpw,Pallis:2018ver} for examples), aiming to introduce no dimensional parameters or explain the heirarchies between different energy scales. As we shall see, whether the scaling symmetry is global or local actually has some important differences. In the presence of a local symmetry, not only a Weyl gauge field has to be accompanying, but also some new term concerning the scalar field appears. Furthermore, the local scaling symmetry indicates the existence of a fundamental energy scale~\cite{Wu:2017urh, Wu:2015wwa}.

This paper is arganized as follows. In Sec.~\ref{sec:formalisim} we discuss the theoretical formalism briefly and illustrate how the classical scaling invariant theory would not be able to give viable inflation. Then in Sec.~\ref{sec:breaking} we show how quantum effects can break the scaling symmetry and induce an effective potential so as to provide viable inflation models. The numerical investigation is presented in Sec.~\ref{sec:nums}, with Fig.~\ref{fig:r-ns} as the main numeric result. Finally, we give our conclusion.

\section{Formalism}\label{sec:formalisim}

The full theory and its formalism has been presented in detail in Ref.~\cite{Wu:2017urh, Wu:2015wwa},  where the gauge-gravity and gravity-geometry correspondences are explicitly demonstrated to obtain the conformal scaling gauge invariant Einstein-Hilbert action for gravitational interaction with a fundamental scalar field~\footnote{One may also look into the nice introductory review~\cite{Hehl:1976kj} for gauge theory of Einstein's gravity.}. For our interest in inflation, we may study the most relevant action with the traditional metric tensor
$g_{\mu\nu}=\chi_{\ \mu}^{a}\chi_{\ \nu}^{b}\eta_{ab}$. The action can be written as follows
\begin{equation}\label{eq:fulllag}
S\equiv \int d^{4}x \mathcal{L}=\int d^{4}x\sqrt{-g}\left[\alpha\left(\phi^{2}R-6\partial_{\mu}\phi\partial^{\mu}\phi\right)+\frac{1}{2}D_{\mu}\phi D^{\mu}\phi-\frac{\beta}{4!}\phi^{4}-\frac{1}{4}W_{\mu\nu}W^{\mu\nu} \right],
\end{equation}
where $g\equiv \textrm{det}\left(g_{\mu\nu}\right)$, $\alpha$ and $\beta$ are constant parameters that will be decided by observations, and covariant derivative $D_\mu $ for real scalar field $\phi$ is given by $D_{\mu}\phi=\left(\partial_{\mu}-g_{W}W_{\mu}\right)\phi$ (note that there is no factor $i$ in front of $g_W$, different from usual $U(1)$ theory),
$g_W$ is a coupling associated with Weyl gauge field $W_{\mu}$. Note that the action is just a subset of the whole action which should include other matter fields (such as fields in standard model) whose effects will be discussed shortly. 

We can also explicitly add a term $\delta \mathcal{L}$ that does not share the same symmetries as other terms. As we will show later, $\delta \mathcal{L}$ is actually crucial to realize a realistic inflation. The above theory has no intrinsic energy scale in the Lagrangian except in $\delta \mathcal{L}$. As long as the dimensional parameters in $\delta \mathcal{L}$ are much smaller than the relevant physical scale, such a theory has the classical scaling symmetry which will be broken by quantum effects. We shall discuss more about this point in Sec.~\ref{sec:breaking}.  

At the moment, let us first ignore $\delta \mathcal{L}$ and focus on the other parts.
Each term in the the rest of the action is invariant under local
conformal scaling symmetry or Weyl symmetry with a positive function $\lambda\left(x\right)$:
\begin{align}\label{eq:conftran}
g_{\mu\nu}\left(x\right) & \rightarrow\bar{g}_{\mu\nu}\left(x\right)=\lambda^{2}\left(x\right)g_{\mu\nu}\left(x\right),
\nonumber \\
\phi\left(x\right) & \rightarrow\bar{\phi}\left(x\right)=\lambda^{-1}\left(x\right)\phi\left(x\right),\nonumber \\
W_{\mu}\left(x\right) & \rightarrow\overline{W}_{\mu}\left(x\right)=W_{\mu}\left(x\right)-\frac{1}{g_{W}}\partial_{\mu}\ln\lambda\left(x\right).
\end{align}
If the above symmetry is exact or when we only consider the classical dynamics, we can easily see that for any generic $g\left(x\right)$ and $\phi\left(x\right)$ it is possible to make a local transformation, by choosing a proper $\lambda\left(x\right)$, to have either $\sqrt{-\bar{g}}=1$
or $\bar{\phi}\left(x\right)=\textrm{const}$, but not simultaneously\footnote{Note that $\left|\bar{g}\right|=1$ does not always mean flat geometry.
	According to Weyl-Schoutem theorem, a Riemannian manifold with dimension
	$n\geq4$ is conformally flat if and only if the Weyl tensor vanishes.
	Nevertheless, for our interest in inflation, we will always focus
	on the case where the metric is conformally flat, $g_{\mu\nu}=a^{2}\left(x\right)\eta_{\mu\nu}$, where $a\left(x\right)$ is the scale factor.}.
If the symmetry was global, $W_\mu$ would be absent. Also the second term in parenthesis with negative sign $-6\partial_{\mu}\phi\partial^{\mu}\phi$ is dropped as in Refs.~\cite{Rinaldi:2014gha, Kannike:2015apa, Ferreira:2018qss, Csaki:2014bua, Kaiser:1994vs}.

Before discussing the theory in Eq.~\ref{eq:fulllag}, we shall warm up with the following action
\begin{equation}\label{eq:simple1}
S=\int d^{4}x\sqrt{-g}\left[\alpha\left(\phi^{2}R-6g^{\mu\nu}\partial_{\mu}\phi\partial_{\nu}\phi\right)-\frac{\beta}{4!}\phi^{4}\right],
\end{equation}
which still preserves the local conformal symmetry
\begin{equation}\label{eq:subtr}
g_{\mu\nu}\left(x\right)\rightarrow\bar{g}_{\mu\nu}\left(x\right)=\lambda^{2}\left(x\right)g_{\mu\nu}\left(x\right),
\phi\left(x\right)\rightarrow\bar{\phi}\left(x\right)=\lambda^{-1}\left(x\right)\phi\left(x\right).
\end{equation}
At first glance, the above theory seems to add a scalar degree of
freedom (dof) to Einstein's general theory of gravity and mimics the scalar-tensor theory. However, if we make a redefinition of metric field
\begin{equation}\label{eq:conf}
\bar{g}_{\mu\nu}\left(x\right)=\Omega^{2}\left(x\right)g_{\mu\nu}\left(x\right),
\end{equation}
where $ \Omega\left(x\right)\equiv\sqrt{2\alpha}\phi/M_{p},\;M_{p}^{2}\equiv 1/(8\pi G)$ is the Planck scale. 
We can rewrite the action with the new field variables
\begin{equation*}
\int d^{4}x\sqrt{-\bar{g}}\Omega^{-4}\left\{ \alpha\left[\phi^{2}\Omega^{2}\left(\bar{R}-6\overline{\square}\ln\Omega+6\bar{g}^{\mu\nu}\partial_{\mu}\ln\Omega\partial_{\nu}\ln\Omega\right)-6g^{\mu\nu}\partial_{\mu}\phi\partial_{\nu}\phi\right]-\frac{\beta}{4!}\phi^{4}\right\} .
\end{equation*}
We have used the relation
\begin{equation*}
R=\Omega^{2}\left[\bar{R}-6\overline{\square}\ln\Omega+6\bar{g}^{\mu\nu}\partial_{\mu}\ln\Omega\partial_{\nu}\ln\Omega\right],\;\overline{\square}\ln\Omega=\frac{1}{\sqrt{-\bar{g}}}\partial_{\mu}\left(\sqrt{-\bar{g}}\bar{g}^{\mu\nu}\partial_{\nu}\ln\Omega\right).
\end{equation*}
The second term in the square bracket will eventually lead to a total
derivative and vanishes on the surface, and the third term cancels with
the derivatives of $\phi$. After substituting $\Omega$, we can obtain
\begin{align}
S & =\int d^{4}x\sqrt{-\bar{g}}\left\{ \left[\frac{M_{p}^{2}}{2}\left(\bar{R}-6\overline{\square}\ln\Omega+6\bar{g}^{\mu\nu}\partial_{\mu}\ln\Omega\partial_{\nu}\ln\Omega\right)-6\alpha\Omega^{-4}g^{\mu\nu}\partial_{\mu}\phi\partial_{\nu}\phi\right]-\frac{\beta}{4!}\frac{M_{p}^{4}}{4\alpha^{2}}\right\} \nonumber\\
 & =\int d^{4}x\sqrt{-\bar{g}}\left[\frac{M_{p}^{2}}{2}\bar{R}-\frac{\beta}{4!}\frac{M_{p}^{4}}{4\alpha^{2}}\right].
\end{align}
This is exactly the Einstein-Hilbert action with a
cosmological constant $\Lambda=\dfrac{\beta}{4!}\dfrac{M_{p}^{4}}{4\alpha^{2}}$
that leads an exponential expansion of Universe for $\beta>0$. This
also shows that total number of physical dofs does not differ from
Einstein's gravity. Extending the discussion into the framework of
quantum field theory does not change the above conclusions since the
number of dofs does not change from classical to quantum theories.
This conclusion is also true even if we break the local scaling symmetry
by adding to the potential with terms that depend on $\phi$ only but not on its derivatives. Adding symmetry-breaking derivative terms would make significant different, as we shall show shortly.

We can get the same result in a different way by choosing $\lambda\left(x\right)=\Omega\left(x\right)$
in Eq.~\ref{eq:subtr}
\begin{align}
\mathcal{L}=\sqrt{-\bar{g}}\left[\alpha\left(\bar{\phi}^{2}\bar{R}-6\bar{g}^{\mu\nu}\partial_{\mu}\bar{\phi}\partial_{\nu}\bar{\phi}\right)-\frac{\beta}{4!}\bar{\phi}^{4}\right] \xrightarrow{\bar{\phi}=\frac{M_{p}}{\sqrt{2\alpha}}} \sqrt{-\bar{g}}\left[\frac{M_{p}^{2}}{2}\bar{R}-\frac{\beta}{4!}\frac{M_{p}^{4}}{4\alpha^{2}}\right],
\end{align}
which effectively chooses a frame $\phi\left(x\right)=M_{p}/\sqrt{2\alpha}$. An interesting observation is that if we choose
$\lambda\left(x\right)=a^{-1}\left(x\right)$($a(x)$ is the scale factor in Friedmann-Walker metric), in such a case we would have the action in a flat spacetime,
\begin{equation}
\bar{R}=0\;\textrm{and }S=\int d^{4}x\left[-6\alpha\eta^{\mu\nu}\partial_{\mu}\bar{\phi}\partial_{\nu}\bar{\phi}-\frac{\beta}{4!}\bar{\phi}^{4}\right].
\end{equation}
The equation of motion for $\bar{\phi}$ is 
\begin{equation}
\eta^{\mu\nu}\partial_{\mu}\partial_{\nu}\bar{\phi}-\frac{\beta}{72\alpha}\bar{\phi}^{3}=0,
\end{equation}
which has a non-trivial solution 
\begin{equation}
\bar{\phi}\left(x\right)=\frac{M}{C\pm x\cdot k},\;\eta^{\mu\nu}k_{\mu}k_{\nu}=\frac{\beta M^{2}}{144\alpha}.
\end{equation}
Here $C$ is an arbitrary constant that should be determined by initial
conditions, $k$ is a constant four-vector and $M$ is an arbitrary energy
scale due to the scaling invariance of this equation. When
we consider homogeneous and isotropic configuration, we have $\bar{\phi}\left(x\right)=M/\left(C\pm x_{0}k_{0}\right)$
which has the same forms as the solutions for exponential expansion
or contraction of the Universe. The above discussions show how equivalent
results can be obtained in two different frames. In the following,
we shall mainly focus on the former or Einstein frame, which has the
advantages when comparing with experiments since essentially all physical
observables are extracted in Einstein frame.

Now we are in the position to discuss the Lagrangian with Weyl gauge field, again ignoring $\delta \mathcal{L}$ at the moment, 
\begin{equation}\label{eq:gaugeL}
\mathcal{L}=\sqrt{-g}\left[\alpha\left(\phi^{2}R-6\partial_{\mu}\phi\partial^{\mu}\phi\right)+\frac{1}{2}D_{\mu}\phi D^{\mu}\phi-\frac{\beta}{4!}\phi^{4}-\frac{1}{4}W_{\mu\nu}W^{\mu\nu}\right].
\end{equation}
In this theory, even if the symmetry is preserved, scalar $\phi$
is a physical degree of freedom. This can be easily shown by replacing
$\phi=v$, the massless gauge boson $W_{\mu}$
gets a mass term from its interaction with $\phi$, $g_{W}^{2}\phi^{2}W_{\mu}W^{\mu}$.
Then $\phi$ is absorbed by $W_{\mu}$ as the longitudial mode, similar
to the Higgs mechanism in usual gauge theories. However,
even if $\phi$ is dynamical, viable inflation is not guaranteed.
Again, if the local conformal symmetry is exact, inflation can occur
but will never stop. To see it explicitly, let us work in the Einstein
frame with $\bar{g}_{\mu\nu}=\Omega^{2}\left(x\right)g_{\mu\nu}$,
\begin{equation}\label{eq:longitude}
S=\int d^{4}x\sqrt{-\bar{g}}\left[\frac{M_{p}^{2}}{2}\bar{R}+\frac{1}{2}\Omega^{-2}\bar{g}^{\mu\nu}D_{\mu}\phi D_{\nu}\phi-\frac{\beta}{4!}\frac{M_{p}^{4}}{4\alpha^{2}}-\frac{1}{4}\bar{g}^{\mu\rho}\bar{g}^{\nu\sigma}W_{\mu\nu}W_{\rho\sigma}\right].
\end{equation}
Rewriting the second term as
\begin{align}
\Omega^{-2}\bar{g}^{\mu\nu}D_{\mu}\phi D_{\nu}\phi & =\frac{M_{p}^{2}}{2\alpha}\bar{g}^{\mu\nu}\left(\frac{1}{\phi^{2}}\partial_{\mu}\phi\partial_{\nu}\phi-\frac{2g_{W}}{\phi}W_{\mu}\partial_{\nu}\phi+g_{W}^{2}W_{\mu}W_{\nu}\right) \nonumber \\
 & =\bar{g}^{\mu\nu}\left(\partial_{\mu}\sigma-m_{W}W_{\mu}\right)\left(\partial_{\nu}\sigma-m_{W}W_{\nu}\right),
\end{align}
where $\sigma\left(x\right)=\text{\ensuremath{\dfrac{M_{p}}{\sqrt{2\alpha}}\ln\dfrac{\phi\left(x\right)}{M_{p}}},\;\ensuremath{m_{W}}=}\dfrac{g_{W}M_{p}}{\sqrt{2\alpha}}.$
Now if we redefine the new gauge field 
\[
\overline{W}_{\mu}=W_{\mu}-\frac{1}{m_{W}}\partial_{\mu}\sigma,
\]
we will get an action without kinetic term for $\sigma$,
\begin{equation}
S=\int d^{4}x\sqrt{-\bar{g}}\left[\frac{M_{p}^{2}}{2}\bar{R}-\frac{\beta}{4!}\frac{M_{p}^{4}}{4\alpha^{2}}-\frac{1}{4}\bar{g}^{\mu\rho}\bar{g}^{\nu\sigma}\overline{W}_{\mu\nu}\overline{W}_{\rho\sigma}+\frac{1}{2}m_{W}^{2}\overline{W}^{\mu}\overline{W}_{\mu}\right].
\end{equation}
The theory now becomes Einstein's gravity with a cosmological constant
and a massive vector field $\overline{W}_{\mu}$. When $\overline{W}_{\mu}$
plays no role in inflation, this theory will have the same difficulty that inflation can start but will not stop. If $\overline{W}_{\mu}$ plays the role as inflaton~\cite{Ford:1989me}, it will suffer instability and turns out to be pathological~\cite{Himmetoglu:2008zp}.

It should be clear by now that the above theories with exact local scaling symmetry would not be able to give realistic inflation models that lead to our present universe. In next section we shall discuss how possible terms might arise in $\delta \mathcal{L}$ can break the local conformal symmetry and lead to viable inflation models.

\section{Breaking of Scaling Invariance by Quantum Effects}\label{sec:breaking}


\begin{figure}[t]
	\includegraphics[width=0.6\textwidth,keepaspectratio]{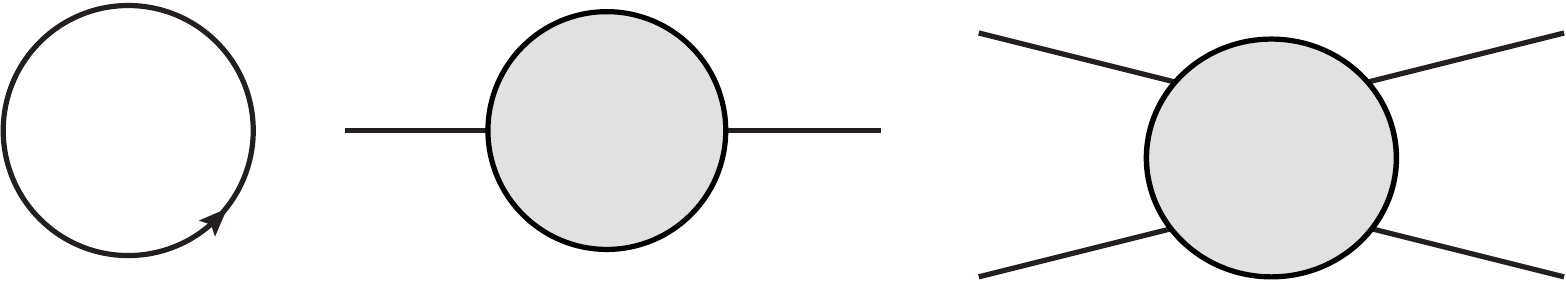}
	\caption{Quantum correction to effective potential. Both $\phi$ and $W_\mu$ can appear in the loop. \label{fig:eff}}
\end{figure}

In this section, we mainly discuss one way to break the scaling symmetry by quantum effects and more options will be briefly outlined in the end of this section. Let us take a step back and work in a familiar and well-understood background, the flat spacetime, the action of Eq.~\ref{eq:gaugeL} is reduced to
\begin{equation}
S=\int d^{4}x\left[-6\alpha \eta^{ab}\partial_{a}\phi\partial_{b}\phi + \frac{1}{2}D_{a}\phi D^{b}\phi-\frac{\beta}{4!}\phi^{4}-\frac{1}{4}W_{ab}W^{ab}\right].
\end{equation}
Strictly speaking, the subscript $a,b=0,1,2,3$ should be interpreted as the index in the local flat frame, within the standard formalism of differential geometry. For our purpose here, without confusion we will use them as same as greek index. 

When quantum corrections are taken into account, something interesting arise from diagrams in Fig.~\ref{fig:eff}, which would contribute to the effective potential $U(\phi)$ for $\phi$.  All of these diagrams are UV divergent, so as standard in quantum field theories, proper counterterms have to be introduced to cancel the divergences. The finite leftover of these cancellations can be determined by renormalization conditions, which equivalently fix the physical input parameters at the renormalization scale $v$. In general, we can write down the finite effective potential $U(\phi)$ as 
\begin{equation}
U(\phi)= A v^4+ B v^2 \phi^2 +  C \phi ^4, 
\end{equation}
where $A, B$ and $C$ are dimensionless functions of $\phi^2/v^2$, whose values are determined by renormalization conditions. For example, in Coleman-Weinberg mechanism~\cite{Coleman:1973jx}, in order to have dimensional transmutation from radiative symmetry breaking, the following conditions are imposed
\begin{equation}
U(0)=0, U'(v)=0, U''(0)=0,
\end{equation}
which lead $A=0,B=0$ and $C\propto \left(\ln \dfrac{\phi^2}{v^2}- \dfrac{1}{2}\right)$. Since our motivation is different from that in Coleman-Weinberg mechanism, we have freedoms in choosing the renormalization conditions for phenomenological studies. Nevertheless, in all cases, it is evidently that introduction of $v$ explicitly breaks the scaling symmetry, which can have significant effects on constructing viable inflation models. Actually, inflation in the original Coleman-Weinberg potential is known to be inconsistent with current observations, which requires modification~\cite{Iso:2014gka} to fit the data.

We give some alternative ways to break scaling symmetry that requires some extra fields. The first one is to introduce a new Yukawa interaction between $\phi$ and some strongly interacting sector, a fermion $\Psi$ with non-Abelian gauge interactions, for instance. Then, terms like $\phi \bar{\Psi}\Psi$ will induce a linear potential for $\phi$ when the confinement happens, $\langle \bar{\Psi}\Psi\rangle \neq 0$. Appearance of linear term $\phi$ can explicitly break the scaling symmetry. Similar mechanism has been used in~\cite{Hur:2011sv, Holthausen:2013ota} to resolve the heirarchy problem in standard model. On the other hand, when the Yukawa interaction becomes strong with a large Yukawa coupling constant, it can also induce a significant quadratic potential term for $\phi$ to cause a dynamical symmetry breaking~\cite{BCW,Wu:2015wwa}.  The second possibility is achieved through an interaction with a new complex scalar field, $S$, with its own gauge interaction. The gauge symmetry can be radiatively broken through Coleman-Weinberg mechanism, then terms like $\phi^2 S^\dagger S$ might trigger the breaking of scaling symmetry in the $\phi$ sector. Models with multiple scalar fields are also discussed in Ref.~\cite{Nishino:2009in, Bars:2013yba}.

\section{Inflation with induced local scaling symmetry breaking}\label{sec:nums}

At this moment, we have all the ingredients to present a realistic inflation model. We use the effective potential $U(\phi)$ and keep $\delta \mathcal{L}$ in Eq.~(\ref{eq:conf})
\begin{equation}\label{eq:inf}
S=\int d^{4}x\sqrt{-g}\left[\alpha\left(\phi^{2}R-6\partial_{\mu}\phi\partial^{\mu}\phi\right)+\frac{1}{2}D_{\mu}\phi D^{\mu}\phi -U(\phi) -\frac{1}{4}W_{\mu\nu}W^{\mu\nu} + \delta \mathcal{L} \right],
\end{equation}
$\delta \mathcal{L}$ will include terms such that redefinition of fields does not make $\phi$ undynamical. Therefore, $\phi$ might be responsible for the inflation. A physical condition for the effective potential $U(\phi)$ is that $U\simeq 0$ at the minimum (or not greater than present dark energy's energy density). For simplicity, at leading order we shall take
\begin{equation}
U(\phi)\propto \bar{\beta} \left(\phi^{2}-v^{2}\right)^{2}.
\end{equation} 
We shall keep in mind that numerical value for $\bar{\beta}$ here could be different from previous $\beta$ since $\bar{\beta}$ is an effective renormalized quantity and runs with renormalization scale. So, even if we start with $\beta=0$ at some scale, quantum corrections will still induce non-vanishing $\bar{\beta}$. This type of potential can be achieved through the following renormalization condition after expanding at $v$,
\begin{equation}
U(v)=0, U'(v)=0, U''(v) \neq 0 .
\end{equation}
Technically speaking, there are logarithmic terms which indicate theoretical parameter's scale-dependence or the RG effects which are usually small for perturbative couplings. As long as $\phi$ is dominating the energy density in the inflationary epoch, we can neglect $W_\mu$'s contribution and work with the following action,
\begin{equation}\label{eq:simple2}
S=\int d^{4}x\sqrt{-g}\left[\alpha\left(\phi^{2}R-6g^{\mu\nu}\partial_{\mu}\phi\partial_{\nu}\phi\right)+\frac{1}{2}g^{\mu\nu}\partial_{\mu}\phi\partial_{\nu}\phi-\frac{\bar{\beta}}{4!}\left(\phi^{2}-v^{2}\right)^{2}\right].
\end{equation}
As already pointed out, this theory does not have the local conformal scaling symmetry due to the presence of $v$. We can still make the same redefinition of the metric field, $g_{\mu\nu}\left(x\right)=\Omega^{-2}\left(x\right)\bar{g}_{\mu\nu}\left(x\right)$, and transform into the Einstein frame,
\begin{align}
	S & =\int d^{4}x\sqrt{-\bar{g}}\left[\frac{M_{p}^{2}}{2}\bar{R}+\frac{1}{2}\Omega^{-2}\bar{g}^{\mu\nu}\partial_{\mu}\phi\partial_{\nu}\phi-\frac{\bar{\beta}}{4!}\frac{M_{p}^{4}}{4\alpha^{2}}\left(1-\frac{v^{2}}{\phi^{2}}\right)^{2}\right]
	\nonumber\\
	& =\int d^{4}x\sqrt{-\bar{g}}\left\{ \frac{M_{p}^{2}}{2}\bar{R}+\frac{1}{2}\bar{g}^{\mu\nu}\partial_{\mu}\sigma\partial_{\nu}\sigma-\frac{\bar{\beta}}{4!}\frac{M_{p}^{4}}{4\alpha^{2}}\left[1-\frac{v^{2}}{M_{p}^{2}}\exp\left(-\frac{2\sqrt{2\alpha}}{M_{p}}\sigma\right)\right]^{2}\right\} ,
\end{align}
where $\sigma\left(x\right)=\dfrac{M_{p}}{\sqrt{2\alpha}}\ln\dfrac{\phi\left(x\right)}{M_{p}}$.
The theory becomes Einstein's gravity plus a scalar field $\sigma$
with potential
\begin{equation}\label{eq:potentialF}
F\left(\sigma\right)=\frac{\bar{\beta}}{4!}\frac{M_{p}^{4}}{4\alpha^{2}}\left[1-\frac{v^{2}}{M_{p}^{2}}\exp\left(-\frac{2\sqrt{2\alpha}}{M_{p}}\sigma\right)\right]^{2}.
\end{equation}
It is now clear why we need to both introduce the kinetic term and modify the potential. Without the kinetic term, the scalar is not dynamical but just an auxiliary field. While if $v=0$, the scalar $\sigma$ is massless with a flat constant potential and there will be no way to stop once inflation starts. We may also compare the above potential with the one in Starobinsky's inflation model, $F(\sigma)\propto \left[1 - \exp\left(-\sqrt{\frac{2}{3}}\frac{\sigma}{M_p}\right)\right]^2$, which has no free parameter and is less flexible than our model. Predictions of the physical parameters will also be different, as we shall show soon. We have also checked that the inflation potential in this model, Eq.~\ref{eq:potentialF}, has not been collected in {\it Encyclopedia Inflationaris}~\cite{Martin:2013tda}. Meanwhile, we note that the model with global scale invariance (without $6\partial_{\mu}\phi\partial^{\mu}\phi$ in the action) would have the following potential that can be computed in similar spirit,
\begin{equation}\label{eq:globalF}
F(\sigma)\propto \left[1-\frac{v^{2}}{M_{p}^{2}}\exp\left(-\frac{2\sigma}{M_{p}}\sqrt{\frac{2\alpha }{1+12\alpha}}\right)\right]^{2}.
\end{equation}
It goes into Eq.~\ref{eq:potentialF} only in the limit of $12\alpha \ll 1$.

\begin{figure}
	\includegraphics[width=0.52\textwidth,height=0.42\textwidth]{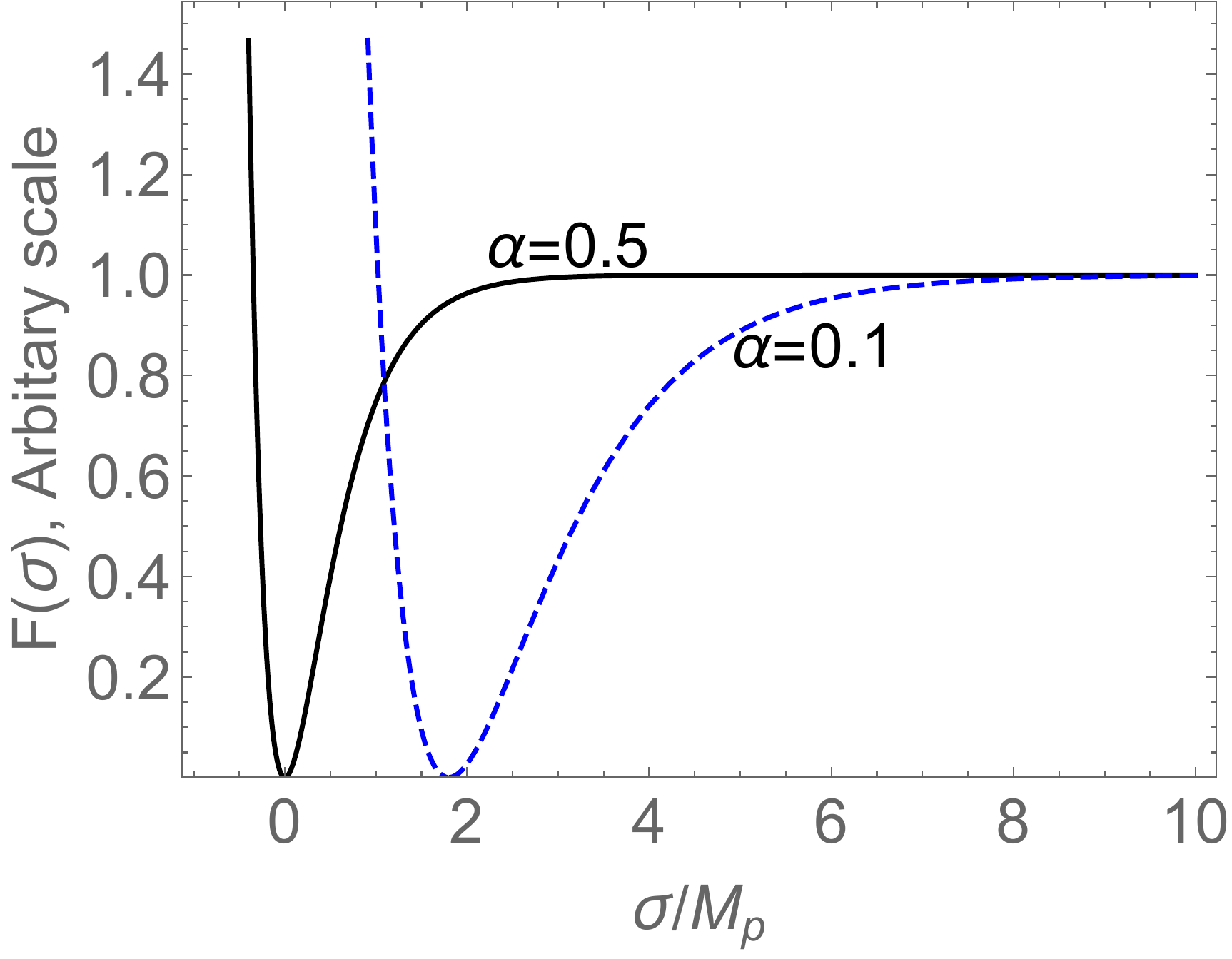}
	\caption{Shape of the potential $F(\sigma)$ in Eq.~\ref{eq:potentialF} for $\alpha=0.5$(solid) and $\alpha=0.1$(dashed) while keeping $\alpha v^2=M^2_p/2$. Flatness or slow-roll conditions can be respected as long as $\sigma$ is far away from the minimum point.  \label{fig:potential}}
\end{figure}

The shape of the potential $F(\sigma )$ in Eq.~\ref{eq:potentialF} is shown in the Fig.~\ref{fig:potential} for $\alpha=0.5$(solid) and $\alpha=0.1$(dashed) while keeping $\alpha v^2=M^2_p/2$, and an overall factor has been normalized to 1 since the shape has no dependence on it, i.e., 
\begin{equation}
F\left(\sigma\right)/ \frac{\bar{\beta}}{4!}\frac{M_{p}^{4}}{4\alpha^{2}} = \left[1-\frac{1}{2\alpha}\exp\left(-\frac{2\sqrt{2\alpha}}{M_{p}}\sigma \right)\right]^{2}. \nonumber 
\end{equation}
The minimum of the potential is reached at 
\begin{equation}\label{eq:min}
\sigma_0=\dfrac{M_{p}}{\sqrt{2\alpha}}\ln\dfrac{1}{\sqrt{2\alpha}}.
\end{equation}
The physical mass of $\sigma$ around $\sigma_0$ is given by
\begin{equation}\label{eq:smass}
m_\sigma = \sqrt{\frac{\bar{\beta}}{12\alpha}}M_p. 
\end{equation}

Now we can use the standard slow-roll inflation formalism. The slow-roll parameters are calculated as
\begin{align}\label{eq:slowroll}
	\epsilon & \equiv\frac{1}{2}M_{p}^{2}\left(\frac{F_{\sigma}\left(\sigma\right)}{F\left(\sigma\right)}\right)^{2}=\frac{16\alpha v^{4}}{\left[v^{2}-\phi^{2}\right]^{2}},\\
	\eta & \equiv M_{p}^{2}\frac{F_{\sigma\sigma}}{F\left(\sigma\right)}=\frac{16\alpha v^{2}\left[2v^{2}-\phi^{2}\right]}{\left[v^{2}-\phi^{2}\right]^{2}},\\
	\xi^{2} & \equiv M_{p}^{4}\frac{F_{\sigma}F_{\sigma\sigma\sigma}}{F^{2}}=\frac{256\alpha^{2}v^{4}\left[4v^{2}-\phi^{2}\right]}{\left[v^{2}-\phi^{2}\right]^{3}},
\end{align}
where we has used $\phi=M_{p}\exp\left(\frac{\sqrt{2\alpha}}{M_{p}}\sigma\right)$. As long as the slow-roll conditions $\epsilon\ll 1$ and $|\eta| \ll 1$ are satisfied, the Universe is undergoing a quasi-exponential expansion. When slow-roll conditions are violated, the Universe will exit from inflation. Then the inflaton $\sigma$ will oscillate around the minimum and decay into other light particles to reheat the Universe. These parameters are related with physical observables, scalar spectral index for the scalar power spectrum
\begin{align}
	n_{s} & =1-6\epsilon+2\eta = 1 - \frac{32\alpha v^2 \left[v^{2}+\phi^{2}\right]}{\left[v^{2}-\phi^{2}\right]^{2}},
\end{align}
and tensor-to-scalar ratio $ r=16\epsilon$. Both are constrained by the recent Planck results~\cite{Ade:2015lrj}, $n_{s} = 0.968\pm 0.006$ and $r \lesssim 0.12$. The running of scalar spectral index is small and given by at second-order,
\begin{equation}
\frac{d n_s}{d\ln k}\simeq -8\epsilon^2(3\epsilon-2\eta)-2\xi^2=-\frac{512 \alpha ^2 v^4 \phi^2 \left(3 v^2+\phi^2\right)}{\left(v^2-\phi^2\right)^4},
\end{equation}
where $k$ is the wave-number. Current observation gives $d n_s/d\ln k=-0.003\pm 0.007$~\cite{Ade:2015lrj}, consistent with zero. The overall amplitudes of scalar and tensor power spectrum are expressed as
\begin{equation}\label{eq:fluct}
\Delta_{s}^{2}\left(k\right)\approx\left.\frac{1}{24\pi^{2}}\frac{F}{M_{p}^{4}}\frac{1}{\epsilon}\right|_{k=aH},\;\Delta_{t}^{2}\left(k\right)\approx\left.\frac{2}{3\pi^{2}}\frac{F}{M_{p}^{4}}\right|_{k=aH},
\end{equation}
where the pivot wave-number $k$ is usually taken at $k=0.05\textrm{Mpc}^{-1},$
and $\Delta_{s}^{2}\left(k\right)\sim2.2\times10^{-9}$ as measured in Planck~\cite{Ade:2015lrj}. Also, the e-folding number $N$,
\begin{equation}
N\equiv \ln \frac{a(t_\textrm{end})}{a(t)}=\int^{t_\textrm{end}}_t Hdt \simeq \frac{1}{M_p}\int ^{\sigma}_{\sigma_\textrm{end}}\frac{d\sigma}{\sqrt{2\epsilon}}=\frac{1}{\sqrt{2\alpha}}\int ^{\phi}_{\phi_\textrm{end}}\frac{d\phi}{\sqrt{2\epsilon}\phi},
\end{equation}
should be comparable to $\sim 60$ in order to solve the flatness problem.

\begin{figure}
	\includegraphics[width=0.65\textwidth,height=0.52\textwidth]{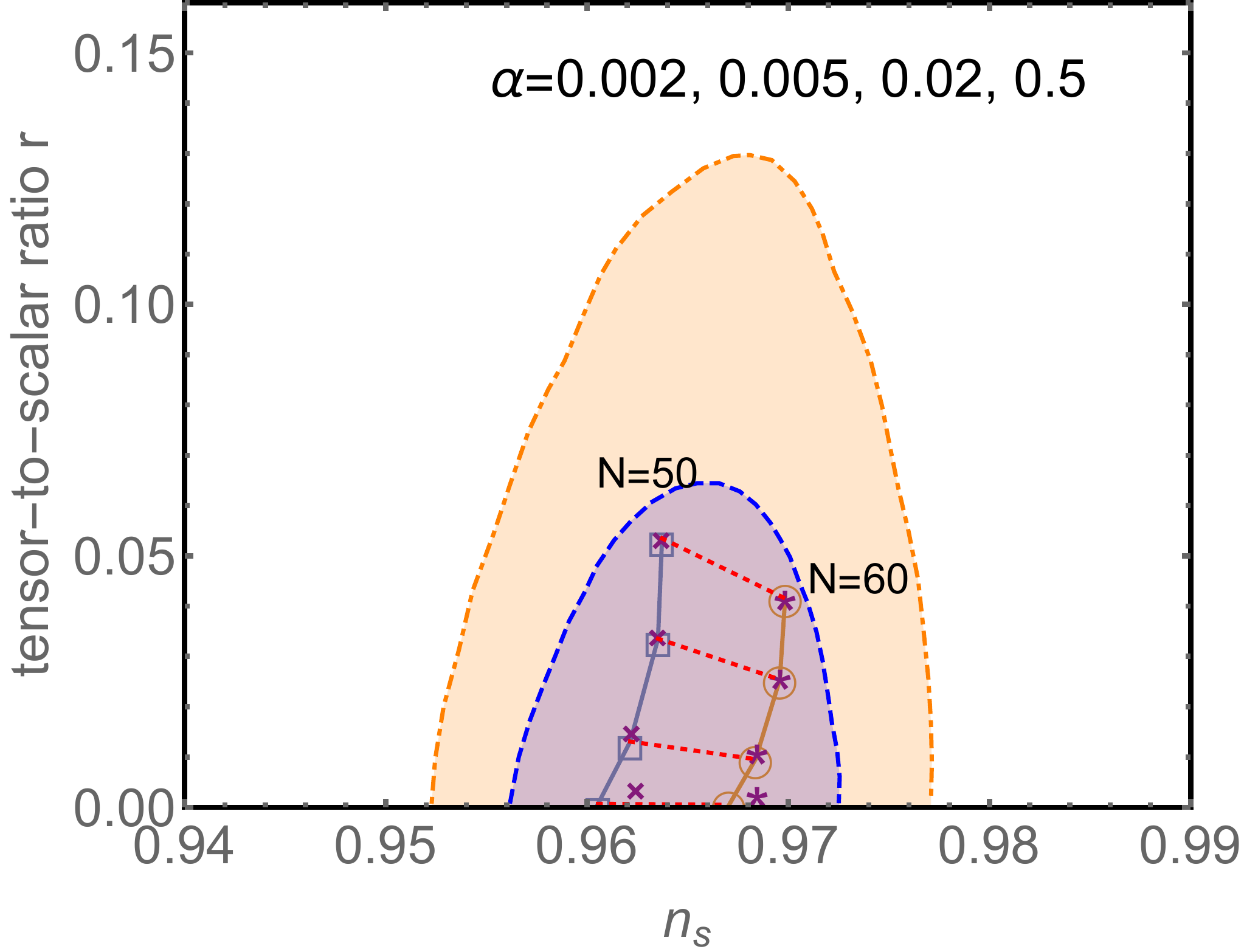}
	\caption{Comparison with observations for $\alpha=0.002,0.005,0.02$ and $0.5$ (from top to bottom). The predictions for $(n_s, r)$ are shown for e-folding number $N=50$(square) and $60$(circle), in comparison with the shaded regions favored by Planck~\cite{Ade:2015lrj} with 1-$\sigma$ (blue) and 2-$\sigma$ (orange). Dashed red lines show the gradual changes from $N=50$ to $N=60$. In all cases, the running index $d n_s/d\ln k\simeq -5\times 10^{-4}.$ We also show $(n_s, r)$ for the same $\alpha$ in models with global scaling symmetry whose potential is given by Eq.~\ref{eq:globalF}, marked by cross and star symbols.
		\label{fig:r-ns}}
\end{figure}

Although inflation could happen at both regions $\phi<v$ or $\phi>v$, cosmological observations has selected the regime $\phi>v$. This can be easily seen from the slow-roll parameter $\eta\simeq2\epsilon$ for $\phi\ll v$, which leads $n_{s}=1-2\epsilon$ or $n_{s}=1-r/8$. Since Planck gives limit $r\lesssim 0.12$, it would give $n_{s}\gtrsim 0.985$ in this region, disfavored by the observation limit $n_{s}=0.968\pm 0.006$. So only $\phi>v$ might be able to give viable models. In such a case, slow-roll conditions are respected as long as $\phi^2>\phi^2_{\textrm{end}}\equiv\left(1+4\sqrt{\alpha}\right)v^2$. 

We numerically compute the corresponding scalar spectral index and tensor-to-scalar ratio with different e-folding number $N=50$ and $60$, and illustrate the theoretical predictions in cases with different $\alpha$, $\alpha=0.002,0.005,0.02$ and $0.5$ (from top to bottom) in Fig.~\ref{fig:r-ns}. The results are shown with different markers for $N=50$(square) and $60$(circle), in comparison with shaded regions favored by Planck~\cite{Ade:2015lrj} with 1-$\sigma$ (blue) and 2-$\sigma$ (orange). In all cases, the running index is very small, $d n_s/d\ln k\simeq -5\times 10^{-4}$. We also show with dashed red lines the gradual changes when e-folding number $N$ goes from $50$ to $60$. It is clearly seen that the theory can fit the observation quite well and fall into the 1-$\sigma$ region. Part of the parameter region will be probed by future cosmological experiments, especially those searching for primordial gravitational wave. In comparison, Starobinsky's inflation will give  $n_s\approx1-2/N\sim 0.967 $ and $r\approx 12/N^2\sim 0.003$ for $N=60$. We also calculate $n_s$ and $r$ for the same $\alpha$ in models with global scaling symmetry whose potential is given by Eq.~\ref{eq:globalF}, marked by cross and star symbols. It explicitly shows that the difference gets smaller as $\alpha$ approaches to 0.

From the amplitude of scalar power spectrum, we can estimate the quartic coupling in the effective potential $\bar{\beta}\sim 1.2\alpha \times 10^{-9}$, which is typical in inflationary models since it is basically the inferred primordial density fluctuation $\sim 10^{-5}$ that determines the overall height of the potential, see Eq.~\ref{eq:fluct} for the exact relation. In a sense, there is only one free parameter in this inflation model, $\alpha$, since others are entirely determined by current observations.

\section{Conclusion}
Motivated by the gauge theory of gravity with local conformal scaling symmetry proposed recently by one of the authors, Wu~\cite{Wu:2017urh, Wu:2015wwa}, we have investigated the possible inflationary dynamics of the scalar field therein. We have shown that the classical theory would not be able to give viable inflation due to the local scaling invariance which dictates an eternal exponential expansion. However, once quantum effects are taken into account, the effective potential can have phenomenologically interesting forms that lead to viable inflation models. 

Main of our numerical results are illustrated in Fig.~\ref{fig:r-ns}, which shows that the theoretical predictions of spectral index and scalar-to-tensor ratio in this modela agree with current observations within $1$-$\sigma$. These parameter space will also be partially probed by future cosmological experiments, such as those searching for primordial gravitational waves.

\section*{Acknowledgments}

YT is supported by the Grant-in-Aid for Innovative Areas No.16H06490. YLW is supported in part by the National Science Foundation of China (NSFC) under Grants \#No. 11690022, No.11475237, No.~11121064,  No. 11747601, and by the Strategic Priority Research Program of the Chinese Academy of Sciences, Grant No. XDB23030100, and by the CAS Center for Excellence in Particle Physics.


\providecommand{\href}[2]{#2}\begingroup\raggedright
\endgroup

\end{document}